\documentclass[10pt,letterpaper]{article}
\usepackage{acl2015,indentfirst,times,secdot}
\usepackage{xspace,empheq,fancybox,amsmath,bbm,amssymb,epsfig,subfig,syntonly,times,amsthm,graphicx} \usepackage{psfrag,color,bm,array}
\usepackage{url,cite,footnote,syntonly,algpseudocode}
\usepackage{booktabs,tabularx}
\usepackage{verbatim,slashbox}
\usepackage[linesnumbered,ruled,vlined]{algorithm2e}

\usepackage{verbatim,multirow,diagbox,enumitem,tabu,acl2015}
\usepackage[T1]{fontenc}
\usepackage{setspace}
\usepackage[justification=justified,font={small,sf}]{caption}
\usepackage[table,xcdraw]{xcolor}
\usepackage{adjustbox,balance}

\usepackage[colorinlistoftodos]{todonotes}

\def\BibTeX{{\rm B\kern-.05em{\sc i\kern-.025em b}\kern-.08em
    T\kern-.1667em\lower.7ex\hbox{E}\kern-.125emX}}

\newtheorem{remark}{\bfseries Remark}
\input{mysymbol.sty}
\graphicspath{{/}}

\newcommand{\edit}{\color{black}{}} 

\begin{document}
    
\title{A Unified Approach for Learning the Dynamics of \\ Power System Generators and Inverter-based Resources}


\author{ \normalfont Shaohui Liu, Weiqian Cai, Hao Zhu, Brian Johnson \\
 Chandra Family Department of Electrical and Computer Engineering \\
 The University of Texas at Austin \\
 {\underline{ \{shaohui.liu, stratoscwq, haozhu, b.johnson\}@utexas.edu}} \\ 
}

 \maketitle


\begin{abstract}
The growing prevalence of inverter-based resources (IBRs) for renewable energy integration and electrification greatly challenges power system dynamic analysis. To account for both synchronous generators (SGs) and IBRs, this work presents an approach for learning the model of an individual dynamic component. The recurrent neural network (RNN) model is used to match the recursive structure in predicting the key dynamical states of a component from its terminal bus voltage and set-point input. To deal with the fast transients especially due to IBRs, we develop a Stable Integral (SI-)RNN to mimic high-order integral methods that can enhance the stability and accuracy for the dynamic learning task. 
We demonstrate that the proposed SI-RNN model not only can successfully predict the component's dynamic behaviors, but also offers the possibility of efficiently computing the dynamic sensitivity relative to a set-point change. These capabilities have been numerically validated based on full-order Electromagnetic Transient (EMT) simulations on a small test system with both SGs and IBRs, particularly for predicting the dynamics of grid-forming inverters. 
\end{abstract}


\section{Introduction}\label{sec:intro}


The electric power infrastructure is witnessing transformative changes with the rapid electrification of various sectors and the increasing installation of renewable energy sources. These changes  have led to growing prevalence of power electronics throughout the power grid~\cite{tolbert2005power}. While traditional rotating machines will continue to play a significant role in generation, storage, and motor-driven applications, the growing role of inverter-based resources (IBRs) presents significant challenges for grid operation and stability. In particular, the presence of IBRs increases the presence of fast transients and increases the complexity of power system dynamics~\cite{taylor2016power}. While Electromagnetic Transient (EMT) simulations could  offer greater precision over the conventional phasor-domain simulations, they are known to be computationally intensive and lack numerical scalability~\cite{nerc2023emt}. Developing effective yet tractable dynamic models for IBRs for system-level simulations is of high importance to facilitate the dynamic studies and stability analysis of power systems.

Recent  advances in data-driven approaches,  especially deep learning techniques based on the neural network (NN) models, have been increasingly utilized for power system dynamical modeling; see e.g.,~\cite{zhao2020overview,huang2023data,roberts2022continuous}. Enabled by the proliferation of sensing infrastructure, the ability to learn dynamics from real data is now feasible. With this approach, NNs offer high model accuracy and are generalizable to different operating conditions.  
However, a majority of these approaches aim to model the full system model along with all of its dynamical subsystems. Such methods incur very high model complexity and limit flexibility under various operating conditions seen in practice. A modularized design where each synchronous generator (SG) has its own dynamical model can be more adaptive and scalable~\cite{li2023integrating,xiao2022feasibility}. 
Despite this advancement, many existing models predict only fixed types of SG dynamic states, overlooking the underlying variations in dynamic models, which are not suitable for learning fast dynamics of IBRs.
An alternative method involves neural ordinary differential equation (neural-ODE) based approaches~\cite{chen2018neural}, which parameterize the state-space model using a neural network, rather than directly predicting the states themselves~\cite{moya2023deeponet,zhanglearning,chevalier2022accelerating}. While this approach is flexible in evaluation points, it necessitates the use of a differential equation solver to generate state predictions. This requirement not only adds a significant computational burden but also complicates system integration if modularized design is considered, particularly with the additional necessity of solving power flow equations.


The goal of our work is to put forth a unified approach for learning the models of dynamical components including both SGs and IBRs. To encompass high-order heterogeneous component models, our method builds upon the actual dynamic models to extract the key state variables that are sufficient for representing the system-level behaviors of these different resources. This reduced-order model is helpful for lowering data rates and model complexity, and allows for the component to interact with the terminal bus only and decouple from the rest of system. The latter is very useful for modular integration of network-wide resources for attaining a full system model. It also allows us to build the learning model based upon the recurrent NN (RNN) to match the dynamic recursion form of the underlying physics-based model. 

Nonetheless, the extremely fast dynamics of IBRs would require the RNN to have a very small time-step and thus high computational complexity. This issue is similar to the stability concern faced by classical numerical integration methods,  as the prediction error per time step can get aggravated through the iterative propagation process. 
To tackle this issue, we put forth an innovative RNN design that mimics the higher-order numerical integral methods which are known to have much better stability guarantees than the first-order Euler's method. We propose the Stable Integral RNN (SI-RNN) model by introducing higher-order blocks of RNNs that are similar to the Runge-Kutta (RK) method~\cite[Ch.~9.3]{heath2018scientific}. Our proposed SI-RNN can use a reasonable time-step to track IBR transients and  greatly improve the stability regions, a crucial consideration in grid dynamic modeling~\cite{ellinas2024correctness}. We have further  leveraged the SI-RNN model towards effectively predicting the dynamic sensitivity with respect to (wrt) a change of  control set-point. This capability is useful for enhancing the generalizability of the trained SI-RNN model,  but more importantly, it  opens up new possibilities for applying to sensitivity-based  optimization and control.

The main contributions of this research are two-fold. First, we present a unified approach to learn the dynamic model of an individual dynamic component like SGs and IBRs. Through reduced-order modeling and network decoupling, our approach can achieve flexible and efficient integration of all network-wide components in a modular fashion.  
Second, we develop a novel RNN-based learning method that can effectively capture the fast transients of IBRs. 
Inspired by the excellent stability guarantees of  high-order integral schemes, our proposed SI-RNN can  significantly reduce the time-step requirement and computation complexity. We demonstrate that this model is not only accurate in predicting the component's dynamics, but also capable of tracking the dynamic sensitivity to set-point changes, thus providing an effective tool for sensitivity-based optimal control tasks.

The rest of the paper is organized as follows. Section~\ref{sec:ps} presents the dynamic models of the SGs and  grid-forming (GFM) IBRs. Section~\ref{sec:learn_model} develops a RNN-based model to learn the component-level dynamics, as well as to predict the dynamic sensitivity. Section~\ref{sec:numerical_results} demonstrates the validity and advantages of our proposed model over the regular RNN model, especially for predicting GFM dynamics. Conclusions and ongoing works are summarized in Section~\ref{sec:con}.
 



\section{Dynamics Modeling}
\label{sec:ps}
We first present the modeling of two types of important power system dynamic components, namely the synchronous generators (SGs) and grid-forming (GFM) inverters. 
Typically, the system dynamics are represented by a set of differential algebraic equations (DAEs), where the differential equations model the dynamics of the component state variables, and algebraic equations capture the power balance; see e.g.,~\cite[Ch.~6-9]{arthur2000power}. {\edit Note that both SGs and IBRs have very high-order dynamics in practice. Nonetheless, we will focus on the low-complexity models of a reduced order that could well represent their behaviors at the power system level. }


\subsection{Modeling of Synchronous Generators (SGs)}

{\edit We consider the the two-axis model of a synchronous machine as described in~\cite[Ch.~5]{sauer_book}, which will be used to incorporate the governor/exciter controls later on.} There are four dynamic states for the SGs: the rotor angle $\delta$ and its derivative (frequency) $\omega$, as well as the d/q-axis internal voltages $E_\mathrm{q}^{\prime}$ and $E_\mathrm{d}^{\prime}$. Using the d/q-axis current magnitudes, $I_\mathrm{d}$  and $I_\mathrm{q}$, per the power flow coupling, we obtain  
\begin{subequations}
\label{eq:sm_dynamics}
\begin{align}
\dfrac{\mathrm{d} \delta}{\mathrm{d} t} & =\omega-\omega_\mathrm{o}, \\
M \dfrac{\mathrm{d} \omega}{\mathrm{d} t} & =P_\mathrm{M}-E_\mathrm{d}^{\prime} I_\mathrm{d}-E_\mathrm{q}^{\prime} I_\mathrm{q} \notag\\
&-(X_\mathrm{q}^{\prime}-X_\mathrm{d}^{\prime}) I_\mathrm{d} I_\mathrm{q}-D(\omega - \omega_\mathrm{o}), \\
T_\mathrm{d o}^{\prime} \dfrac{\mathrm{d} E_\mathrm{q}^{\prime}}{\mathrm{d} t} & =-E_\mathrm{q}^{\prime}-\left(X_\mathrm{d}-X_\mathrm{d}^{\prime}\right) I_\mathrm{d}+E_\mathrm{f d} ,\\
T_\mathrm{q o}^{\prime} \dfrac{\mathrm{d} E_\mathrm{d}^{\prime}}{\mathrm{d} t} & =-E_\mathrm{d}^{\prime}+\left(X_\mathrm{q}-X_\mathrm{q}^{\prime}\right) I_\mathrm{q}, 
\end{align}
\end{subequations}
where 
$\omega_\mathrm{o}$ stands for the nominal frequency (e.g., the synchronous frequency), while 
the field voltage $E_\mathrm{f d}$ and the mechanical power $P_\mathrm{M}$ are determined by the SG controllers. The other parameters in \eqref{eq:sm_dynamics} are mostly fixed constants for each SG, including inertial, damping, and reactance constants.  {\edit Note that this fourth-order model in \eqref{eq:sm_dynamics} is a simplified approximation of the higher-order dynamics including additional state variables like sub-transient voltages and fluxes. }

The SG internal controllers include the excitation systems and governors~\cite[Ch.~8.6]{sauer_book}. 
These controllers typically follow the proportional-integral control rule based on their corresponding set-points. {\edit Since the controller states are typically unobservable for SGs, we only consider simplified state-space models for SG controller states that have direct impacts on SG internal states.} Specifically, the exciter control using the voltage reference set-point signal $V^\star$  can be  represented by:
\begin{align}
T_\mathrm{A} \dfrac{\mathrm{d}{E}_\mathrm{fd}}{\mathrm{d}t} = -{E}_\mathrm{fd} + K_\mathrm{A} (V^\star - V) \;,
\label{eq:exciter_linear}
\end{align}
where $V$ stands for the bus voltage magnitude measured at the generator terminal.
Similarly for the governor, 
it can be simplified as a first-order system using the power set-point signal $P^\star$:
\begin{align}
T_\mathrm{S V} \dfrac{\mathrm{d} P_\mathrm{M}}{\mathrm{d} t} & =-P_\mathrm{M}+P^\star-\dfrac{1}{R_\mathrm{D}}\left(\dfrac{\omega}{\omega_\mathrm{o}}-1\right) \;.
\label{eq:governor_linear}
\end{align}

To determine the currents $I_\mathrm{d}$ and $I_\mathrm{q}$, we solve the power flow equations between the SG's interval voltages and its terminal bus voltage phasor, $V\angle\theta$, given by
\begin{subequations}
\begin{align}
\label{eq:SG_pf}
R I_\mathrm{d}-X_\mathrm{q} I_\mathrm{q}&=E_\mathrm{d}^{\prime}-V \sin (\delta-\theta), \\
R I_\mathrm{q}+X_\mathrm{d} I_\mathrm{d}&=E_\mathrm{q}^{\prime}-V \cos (\delta-\theta) \;.
\end{align}
\end{subequations}

The SG states can be fully determined by the control signals $V^\star$ and $P^\star$, as well as the terminal bus voltage $V\angle\theta$. If the control signals and terminal voltages are all given, the SG dynamics is decoupled from the remaining system. This will be used as the basis for component-wise dynamics learning. 

\subsection{Modeling of Inverter-based Resources (IBRs)}
\begin{figure}[t!]
\centering
\includegraphics[width=1\linewidth]{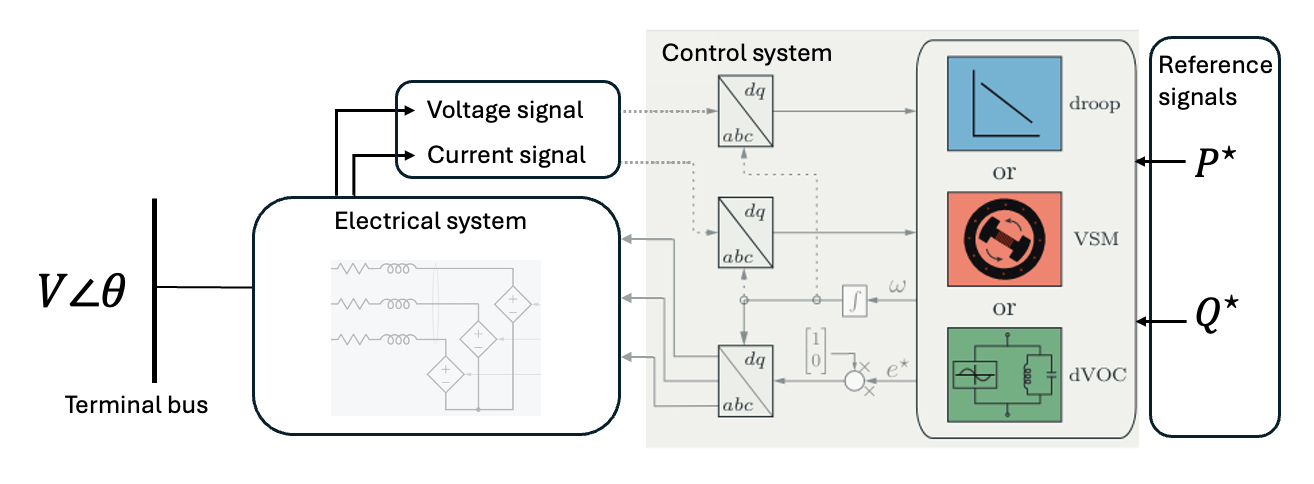}
\caption{Schematic diagram of the averaged full-order GFM-inverter model with the reference signals, control system, and electrical system~\cite{johnson2022generic}.}
\label{fig:gfm_example}
\end{figure} 

IBRs are broadly categorized into grid-following (GFL) and grid-forming (GFM) inverters based on the control objective design.
While the majority of IBRs currently installed into the power grid are in GFL mode, the GFM inverters are envisioned to be more prevalently used to replace SGs thanks to their capabilities of providing voltage and frequency supports~\cite{lin2020research}. There have also been significant research efforts into developing a unified GFM model with various types of controller designs, as discussed in the recent paper  \cite{johnson2022generic}. 
Thus, this work focuses on the modeling and implementation of GFM inverters. 

Unlike SGs, the dynamics of the GFM inverters are dominated by their primary power control loop instead of physical laws, as depicted in Fig.~\ref{fig:gfm_example}. 
Typical GFM control strategies, including droop, virtual synchronous machine (VSM), and dispatchable virtual oscillator (dVOC), have been unified as a generic primary control model in~\cite{johnson2022generic}. 
{\edit This unification can be further simplified under assumptions of single inductor filter dynamics, pure inductive line impedance, stiff grid, and perfect grid voltage synchronization in the controller. 
There are seven dynamic state variables in the generic GFM model after the simplification}: the phase angle of the inverter internal voltage $\delta$, inverter voltage frequency $\omega$, reference voltage magnitude $E^\star$, filtered real and reactive power $P_\mathrm{m}$ and $Q_\mathrm{m}$, and inverter output current magnitude in synchronous reference frame $I_\mathrm{d}$ and $I_\mathrm{q}$ that rotate at nominal frequency.
Specifically, the generic GFM inverter model can be written as
\begin{subequations}\label{eq:GFMctrl}
\begin{align}
\dfrac{\mathrm d\delta}{\mathrm d t} =& \omega-\omega_\mathrm{o},
 \\
{\tau_{\mathrm f}}\dfrac{\mathrm{d}\omega}{\mathrm{d}t} = & \left({{\kappa}_{\mathrm d}}+1\right)\left(\omega_{\mathrm o}-\omega\right) \nonumber +{\kappa}_{\mathrm f}
\left({P}^{\star} - {P}_{\mathrm{m}}\right), \\
{\tau_{\mathrm v}}\dfrac{\mathrm{d}E^\star}{\mathrm{d}t} = &{f_{\mathrm v}(E^\star)}+ {{\kappa}_{\mathrm v}}\left({Q}^{\star} - {Q}_{\mathrm{m}}\right),\\
\tau_\mathrm{p}\dfrac{\mathrm{d}P_{\mathrm{m}}}{\mathrm{d}t}= &\dfrac{3}{2}VI_\mathrm{d} - {P}_{\mathrm{m}},\\
\tau_\mathrm{p}\dfrac{\mathrm{d}Q_{\mathrm{m}}}{\mathrm{d}t} = &-\dfrac{3}{2}VI_\mathrm{q} - {Q}_{\mathrm{m}},\\
L\dfrac{\mathrm{d}I_{\mathrm{d}}}{\mathrm{d}t} = & E^\star\cos(\delta-\theta)-V + \omega_\mathrm{o} L I_\mathrm{q}, \label{eq:GFM_Id}\\
L\dfrac{\mathrm{d}I_{\mathrm{q}}}{\mathrm{d}t} = &E^\star\sin(\delta-\theta) -\omega_\mathrm{o} L I_\mathrm{d}, \label{eq:GFM_Iq}
\end{align}
\end{subequations}
where $P^\star$ and $Q^\star$ are respectively the active and reactive power reference set-points dispatched by the secondary controllers. Additionally, $\tau_\mathrm{f},\,\tau_\mathrm{v},\,\tau_\mathrm{p},\,\kappa_\mathrm{d},\,\kappa_\mathrm{f},$ and $\kappa_\mathrm{v}$ are fixed
constant parameters and take different forms based on different GFM control strategies, as
detailed in Table~\ref{tab:param}. Note that $f_\mathrm{v}\left(E^\star\right)$ also takes different forms per primary control strategies. While Droop and VSM consider the voltage deviation $V_\mathrm{o}-E^\star$ as $f_\mathrm{v}\left(E^\star\right)$ from nominal voltage $V_\mathrm{o}$, dVOC considers a nonlinear function where
\begin{align*}
   f_\mathrm{v}\left(E^\star\right) = E^\star\left(V_\mathrm{o}^2-E^{\star2}\right) \;. 
\end{align*}


\begin{table}
    \centering
    \begin{tabular}{c|cccccc}
    \toprule
    \midrule
        & $\tau_\mathrm{f}$ & $\tau_\mathrm{v}$ & $\tau_\mathrm{p}$ & $\kappa_\mathrm{d}$ & $\kappa_\mathrm{f}$ & $\kappa_\mathrm{v}$ \\
        \midrule
    Droop & 0 & 0 & $\dfrac{1}{\omega_\mathrm{c}}$
    & 0 & $\dfrac{1}{d_\mathrm{f}}$ & $\dfrac{1}{d_\mathrm{v}}$  \\
        \midrule
    VSM & $\dfrac{m_\mathrm{f}}{d_\mathrm{f}}$ & 0 & $\dfrac{1}{\omega_\mathrm{c}}$
    & $\dfrac{d_\mathrm{d}}{d_\mathrm{f}}$ & $\dfrac{1}{d_\mathrm{f}}$ & $\dfrac{1}{d_\mathrm{v}}$ \\
        \midrule
    dVOC & 0 & $\dfrac{1}{\omega_\mathrm{o}\kappa_2}$ & 0
    & 0 & $\dfrac{\omega_\mathrm{o}\kappa_1}{(E^{\star})^2}$ & $\dfrac{\kappa_1}{\kappa_2E^\star}$ \\
         \midrule
\bottomrule
    \end{tabular}
    \caption{Parametric assumptions under which droop, VSM, and dVOC are transformed into  generic primary control model. The definition of these parameter follows~\cite{johnson2022generic}.}
    \label{tab:param}
\end{table}

\vspace*{2pt}
In general, both the component-level dynamics for SGs~\eqref{eq:sm_dynamics} and GFMs~\eqref{eq:GFMctrl}  can be represented as the first-order ODE system
\begin{align}
\dot{\bbx} = \bbf(\bbx,\bby,\bbu) \;,
\label{eq:general_dynamics}
\end{align}
where the vector $\bbx$ includes the dynamic state variables, vector $\bby$ contains the algebraic variables, while vector $\bbu$ has the control set-points as the input. The vector $\bby =[V,\theta]^\top$ essentially represents the voltage phasor of the terminal bus, which allows to decouple the dynamic component from the rest of network. Note that the current variables are also present in the dynamics, they can be solved from exactly using the internal voltage and terminal voltage, as in \eqref{eq:SG_pf}. In addition, the vector 
$\bbu$ comes from the reference set-point signals determined by the secondary controls. 
{\edit Furthermore, from the perspective of system-level behaviors, certain dynamic states in $\bbx$ can be simplified, as well. Specifically, if we assume GFM currents remain relatively stable, \eqref{eq:GFM_Id}-\eqref{eq:GFM_Iq} are equivalent to \eqref{eq:SG_pf}. Additionally, $E^{'}_\mathrm{d}$ can be eliminated in~\eqref{eq:sm_dynamics}, as $T^{'}_\mathrm{qo}$ is usually sufficiently small.} If we further keep the internal controllers as implicit functions, the dynamic states $\bbx$ in \eqref{eq:general_dynamics} are the trio: internal angle $\delta$, internal frequency $\omega$, and internal voltage magnitude $E$. These simplifications allow to develop a general approach for learning the model of component dynamics as follows.

\begin{remark}[Model simplification]
\label{rmk:reduction}
The simplification of the dynamic states  $\bbx$  is justified by the negligible impact of high-frequency oscillation modes at the system level, as evidenced by our empirical findings from full-state Electromagnetic Transient (EMT) simulations for GFMs. 
{\edit These simulations demonstrate that the high-frequency modes of the internal states are effectively filtered out before influencing system-level dynamics, and confirming the low variability of GFM currents, whose transient disturbances are attenuated by the LCL filters~\cite{resnik2014lcl}. Following this filtering process, our methods consider only state variables in $\bbx = \left[ \delta,\omega,E \right]^\top$ without other dynamics.} Notably, the datasets generated for dynamic learning reflect this smoothed, system-level behavior, enabling simplifications that avoid the unnecessary complexity of transient high-frequency modes in the modeling process, as detailed in Section~\ref{sec:numerical_results}.

\end{remark}


\begin{figure}[t!]
\centering
\includegraphics[width=0.7\linewidth]{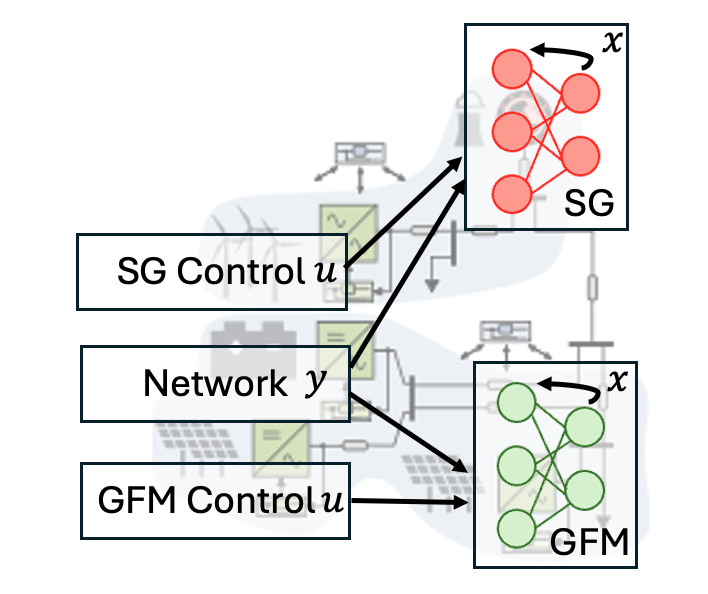}
\caption{Modular architecture for modeling grid dynamics with each component represented as individual NNs. Each component has independent dynamic states, control inputs, and electrical input from the network~\cite{venkatramanan2022integrated}.}
\label{fig:modular_learning}
\end{figure} 


\section{Learning the Component Dynamics}
\label{sec:learn_model}


We consider learning the dynamics of the component system in \eqref{eq:general_dynamics} by directly predicting the state trajectory $\bbx(t)$. Our design falls under the general framework for dynamics prediction in~\cite{li2020machine,li2023integrating}. As discussed later, by building the component-level predictive models, we could flexibly incorporate the system-wide SGs and IBRs into an integrated model, as depicted in Fig.~\ref{fig:modular_learning}. 


To this end, we propose a recurrent neural network (RNN) model with a recursive structure that is nicely suited for predicting the dynamic evolution in \eqref{eq:general_dynamics}. 
Using a sufficiently small time-step $\Delta t$, a simple numerical integration method like  the Euler forward method converts \eqref{eq:general_dynamics} into the discrete-time system
\begin{align}
\bbx_{t+1} \approxeq \bbx_{t} + \Delta t \cdot f(\bbx_t,\bby_t,\bbu_t) \;
\label{eq:fd_euler}
\end{align}
where the approximation accuracy depends on the time-step $\Delta t$, as discussed soon. Without knowing the underlying model~\eqref{eq:general_dynamics}, the integrator in~\eqref{eq:fd_euler} can be approximated using a multi-layer Elman RNN~\cite{elman1990finding}, given by 
\begin{subequations}
\begin{align}
\bbh_t = &\sigma \left(\bbW_{x h}\bbx_t +\bbW_{y h}\bby_{t} \right. \notag\\
&+\bbW_{u h}\bbu_{t}+\bbW_{h h}\bbh_{t-1}+\bbb_{1}\left.\right), \label{eq:rnn_single} \\
\bbx_{t+1} = & \sigma \left(\bbW_{h}\bbh_t +\bbb_{h}\right) \;,
\end{align}
\label{eq:rnn_model}
\end{subequations}
where $\bbh_i$'s are the hidden states in the RNN model. The RNN has trainable parameters denoted by $\bbW_{\cdot},\bbb_{\cdot}$ which are of proper dimensions. For the activation function $\sigma(\cdot)$, typical choices include {\texttt{relu} and \texttt{tanh}}. RNNs are efficient in capturing temporal dependencies through the introduction of the hidden states. Basically, the hidden states act as a form of memory by retaining the important information on the past inputs, allowing the network generates outputs based on both the current input and the information it has captured in the past. 

\begin{figure}[t!]
\centering
\includegraphics[width=1\linewidth]{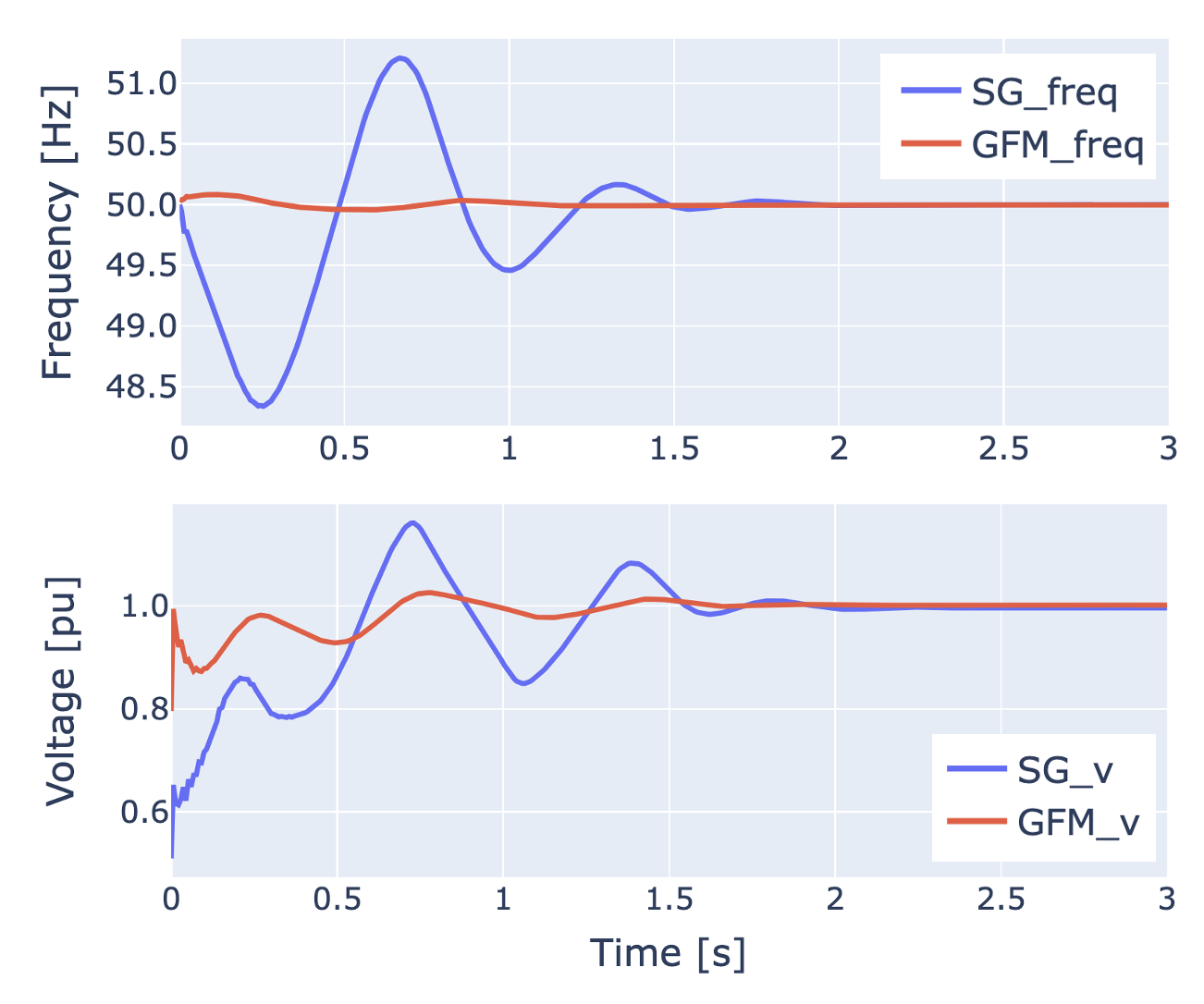}
\caption{Comparison of SG and GFM state responses to a three-phase grounded fault of the 5-bus system in Fig.~\ref{fig:GFM_diagram}.}
\label{fig:gfm_sg_resp}
\end{figure}

While the RNN structure nicely fits with the dynamic learning task, the approximation error introduced by the discretization in \eqref{eq:fd_euler} can adversely affect the accuracy of the trained RNN model. This is related to the well-known stability issue for numerical integration methods based upon \eqref{eq:fd_euler}. 
In order to satisfy the so-termed CFL condition for the stability~\cite[Ch.~11]{heath2018scientific}, the step size $\Delta t$ could be extremely small if \eqref{eq:general_dynamics} is stiff, or equivalently, having very fast dynamics. 
In general, the high-stiffness arises from systems with a high IBR penetration due to the fast dynamics and nonlinear behaviors of IBRs, as shown in Fig.~\ref{fig:gfm_sg_resp}. For dynamical systems, the approximation error of $f(\cdot)$  would accumulate across time, leading to severe error propagation. Thus, it is imperative to improve the RNN design to address the pronounced stability issue with IBR-rich power systems. 

\begin{figure}[t!]
\centering
\includegraphics[width=1.0\linewidth]{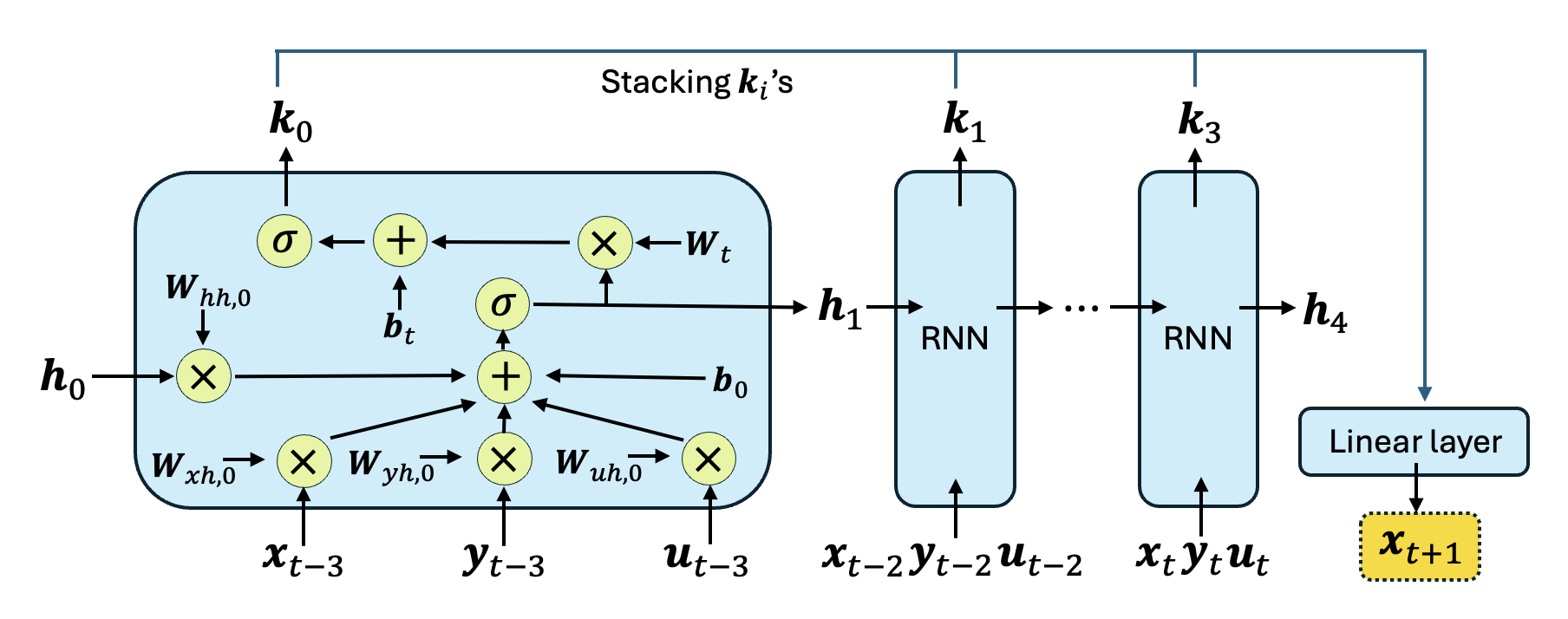}
\caption{Diagram of the proposed SI-RNN dynamics learning framework.}
\label{fig:rnn_diagram}
\end{figure} 

Inspired by higher-order numerical integration methods, we develop a new Stable Integral RNN (SI-RNN) design by imitating the fourth-order explicit Runge-Kutta (RK) method~\cite[Ch.~9.3]{heath2018scientific}. In general, the explicit RK methods use the intermediate steps to reduce the truncation error. For example, consider the fourth-order RK method given by   
\begin{subequations}
    \begin{align}
    \bbx_{t+1} &\approxeq \bbx_t + \Delta t \cdot \sum_{i=0}^3 c_i \bbk_i \,, \\
    \bbk_i &=  f(t+\alpha_i\Delta t,\bbx_t + \Delta t\sum_{j=0}^{i-1} \beta_{i,j}\bbk_j,\bby_t,\bbu_t)
\end{align}
\label{eq:rk_l}
\end{subequations}
where $\{c_i,\alpha_i,\beta_{i,j}\}$ and $\bbk_i$'s are respectively the weight coefficients and functional values for the intermediate steps. 
Compared with the Euler method with an accumulated approximation error in the order of $O(\Delta t)$, the fourth-order RK in \eqref{eq:rk_l} has the total error in the order of $O(\Delta t^4)$. This way, the latter can conveniently handle stiff problems by significantly expanding the stability region for the same time-step. 

To imitate the design \eqref{eq:rk_l}, our proposed SI-RNN leverages multiple historical data points by stacking the recurrent structure in~\eqref{eq:rnn_single}. Fig.~\ref{fig:rnn_diagram} illustrates this idea by stacking  the RNNs for the current input and three previous ones. Each one of the four $\bbk_i$'s is predicted by an individual RNN, with the hidden states $\bbh_i$'s also connected. Thus, our SI-RNN can be described  as
\begin{subequations}
\begin{align}
\bbh_{i+1} = &\sigma \left(\bbW_{x h,i}\bbx_{t-3+i} +\bbW_{y h,i}\bby_{t-3+i}  \right. \notag\\
&+\bbW_{u h,i}\bbu_{t-3+i} +\bbW_{h h,i}\bbh_{i} +\bbb_{1,i}\left.\right) \label{eq:rnn_single_rk} \\
\bbk_{i} = &\sigma \left(\bbW_{h,i}\bbh_{i+1} +\bbb_{i}\right),i=0,\cdots,3 \\
\bbx_{t+1} = &\sigma \left(\bbW_\ell \left[\bbk_0,\cdots,\bbk_3\right]  + \bbb_\ell  \right)\;.
    \label{eq:rnn_rk_linear}
\end{align}
\label{eq:rnn_model_rk}
\end{subequations}
Basically, each of the four RNNs is used to predict the corresponding $\bbk_i$ in the RK method. Note that unlike the typical Elman RNN~\eqref{eq:rnn_model}, the hidden states $\bbh_i$ in \eqref{eq:rnn_model_rk} only track temporal dependencies among the output and four input steps. This special design helps in learning intricate dynamic patterns that might be neglected by the model simplification and state reduction steps without suffering from additional training difficulties, as detailed later in Section~\ref{sec:numerical_results}.
Additionally, the last step in \eqref{eq:rnn_rk_linear} is a linear layer that matches the linear combination in~\eqref{eq:rk_l}. All the weight coefficients $\bbW_\cdot$'s and $\bbb_\cdot$'s are trainable parameters for the SI-RNN. 

To simplify the notations, we represent the SI-RNN model as
\begin{align}
    f_{R} \left( \bbx, \bby,\bbu;\bbTheta  \right),
    \label{eq:rnn_notation}
\end{align}
where $\bbTheta$ contains all trainable parameters in \eqref{eq:rnn_model_rk}. To generate the training datasets, the trajectories of the component states and algebraic variables, along with the set-point signals, are dissected into tuples:
\begin{align*}
     \{ \texttt{'Input:'} \bbX_t,\!\,\texttt{'Output:'} \bbx_{t+1}\}
\end{align*}
with the RNN input
\begin{align*}
    \bbX_t \triangleq \left\{\left\{\bbx_{t-i},\bby_{t-i},\bbu_{t-i}\right\}_{i=0,\cdots,3}\right\}. 
\end{align*}
As for the loss function used by the SI-RNN training, one can compare the predicted $\hhatbbx_{t+1}=f_{R} \left( \bbX_t;\bbTheta \right)$ with actual states to form
\begin{align}
    \ccalL \left( \bbx;\bbTheta \right) \triangleq \frac{1}{T}\sum_{t=1}^T \|\bbx_{t} - \hhatbbx_{t} \|^2_2.
      \label{eq:rnn_loss}
\end{align}

We can use the SI-RNN model developed for an individual dynamic component to come up with an integrated system model of multiple components. 
\begin{remark}[Network integration] 
    The SI-RNN design is convenient for performing an integrated network analysis with multiple, heterogeneous dynamic components. First, the development of the SI-RNN model is applicable to any type of dynamic resources, either SGs or IBRs. Through the setup of $\bby$ and $\bbu$ in Section~\ref{sec:ps}, the component-level model only needs to interface with the rest of power network at the terminal bus, as shown in Fig.~\ref{fig:modular_learning}. This allows for separately training the SI-RNN for any dynamic component, followed up an integration step based on static power flow coupling only. This way, the network model can conveniently adapt to the connectivity of each component, as well as the network connectivity in terms of line status.     
    This flexible integration of network-wide resources is unlike the neural-ODE method~\cite{moya2023approximating,moya2023deeponet} that  predicts  $f(\cdot)$ in \eqref{eq:general_dynamics} as a function of time $t$ and cannot used in the terminal coupling directly. We will further explore this network integration step in future. 
\end{remark}


\subsection{Dynamic Sensitivity}


Our SI-RNN model allows for quickly predicting the dynamic sensitivity with respect to (wrt) the change of control set-points. This is important  for a variety of  power system dynamic analysis tasks. For example, dynamic sensitivity factors can facilitate  the development of measurement-based system identification \cite{zhang2016dependency} and fast contingency screening~\cite{al2019dynamic}. In addition, the predicted state trajectories could be used in the design of optimal control strategies~\cite{gao2022inverse,novak2020supervised}. 

While various kinds of dynamic sensitivity can be considered for the general model \eqref{eq:general_dynamics}, we take the example of the $P^\star - \omega$ sensitivity. This stands for the sensitivity in the frequency $\omega(t)$ wrt a  small change in the active power set-point $P^\star$. If one has the access to the numerical simulation data under different power set-points, the trajectory of this sensitivity can be approximated numerically. Using a small perturbation of $\epsilon$ on the set-point, the finite difference gives
\begin{align}
    J(\omega(t);\Delta P^\star) \triangleq \dfrac{\Delta \omega(t)}{\Delta P^\star} \approx \dfrac{\omega_\epsilon(t)-\omega_0(t)}{\epsilon}\,,
    \label{eq:data_sensitivity}
\end{align}
where $\omega_\epsilon(t)$ and $\omega_0(t)$ are the frequency trajectories under the set-points $P^\star+\epsilon$ and $P^\star$, respectively. Note that when generating the  $\omega(t)$ and $\omega_0(t)$ trajectories, the same initial conditions and set-points are used, except for the change in the set-point $P^\star$.

Since the SI-RNN model has been trained to match the underlying dynamics, it can be used for directly predicting the dynamic sensitivity for a variety of operating conditions. In fact, it  conveniently facilitates this computation thanks to the fast gradient calculation via back-propagation.   
In addition to predicting the trajectory of $\omega(t)$ with given initial conditions and other inputs, the SI-RNN can efficiently produce the sensitivity by computing the partial derivative of $\omega(t)$ wrt $P^\star$, as follows:
\begin{align}
    &\hhatJ(\omega(t);\Delta P^\star) \triangleq \dfrac{\partial\omega(t)}{\partial P^\star} \notag\\
    = &\dfrac{\partial\omega(t)}{\partial \bbk_0}\dfrac{\partial \bbk_0}{\partial \bbh_{1}}\dfrac{\partial \bbh_{1}}{\partial P^\star} \notag\\
    + &\dfrac{\partial\omega(t)}{\partial \bbk_1}\dfrac{\partial \bbk_1}{\partial \bbh_{2}}\left( \dfrac{\partial \bbh_{2}}{\partial P^\star} + \dfrac{\partial \bbh_2}{\partial \bbh_{1}}\dfrac{\partial \bbh_{1}}{\partial P^\star} \right)  \notag\\
    + &\dfrac{\partial\omega(t)}{\partial \bbk_2}\dfrac{\partial \bbk_2}{\partial \bbh_{3}}\left( \dfrac{\partial \bbh_{3}}{\partial P^\star} + \dfrac{\partial \bbh_3}{\partial \bbh_{2}} \left( \dfrac{\partial \bbh_{2}}{\partial P^\star} + \dfrac{\partial \bbh_2}{\partial \bbh_{1}}\dfrac{\partial \bbh_{1}}{\partial P^\star} \right) \right)  \notag\\
    + &\dfrac{\partial\omega(t)}{\partial \bbk_3}\dfrac{\partial \bbk_3}{\partial \bbh_{4}}\left( \dfrac{\partial \bbh_{4}}{\partial P^\star} + \dfrac{\partial \bbh_4}{\partial \bbh_{3}} \left( \dfrac{\partial \bbh_3}{\partial \bbh_{2}} \left( \dfrac{\partial \bbh_{2}}{\partial P^\star} + \dfrac{\partial \bbh_2}{\partial \bbh_{1}}\dfrac{\partial \bbh_{1}}{\partial P^\star} \right) \right) \right)  \notag\\
    \label{eq:rnn_sensitivity}
\end{align}
where the second step follows from the chain rule, and can be efficiently calculated through recursion.

\begin{remark}[Sensitivity-aware learning]
The SI-RNN model-based sensitivity \eqref{eq:rnn_sensitivity} can be efficiently computed using the automatic differentiation and reverse accumulation in Pytorch, as implemented  by the widely adopted backpropagation method in NN training~\cite{werbos1982applications}. 
This approach not only alleviates the need of simulating multiple trajectories in the numerical estimation approach based on \eqref{eq:data_sensitivity}, but also can be utilized for training the SI-RNN model to match a given sensitivity trajectory. By regularizing the SI-RNN loss function in \eqref{eq:rnn_loss} with the mismatch between the predicted and actual sensitivity, we can further improve the model accuracy  
and generalizability performance.  
\end{remark}



\section{Numerical Results}
\label{sec:numerical_results}

\begin{figure}[t!]
\centering
\includegraphics[width=1\linewidth]{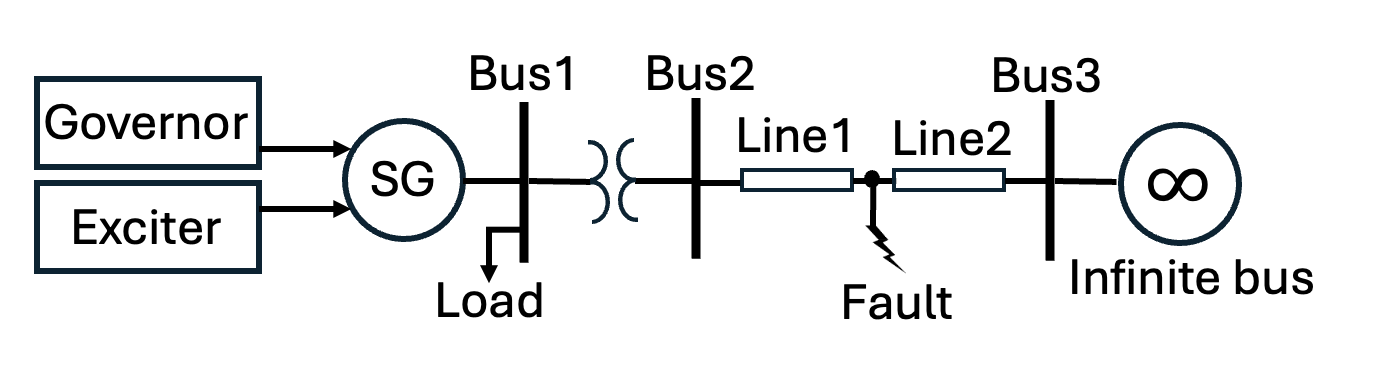}
\caption{ One-line diagram of the SMIB system for learning the SG dynamics.}
\label{fig:SMIB_diagram}
\end{figure} 

\begin{figure*}[t!]
\centering
\includegraphics[width=1\linewidth]{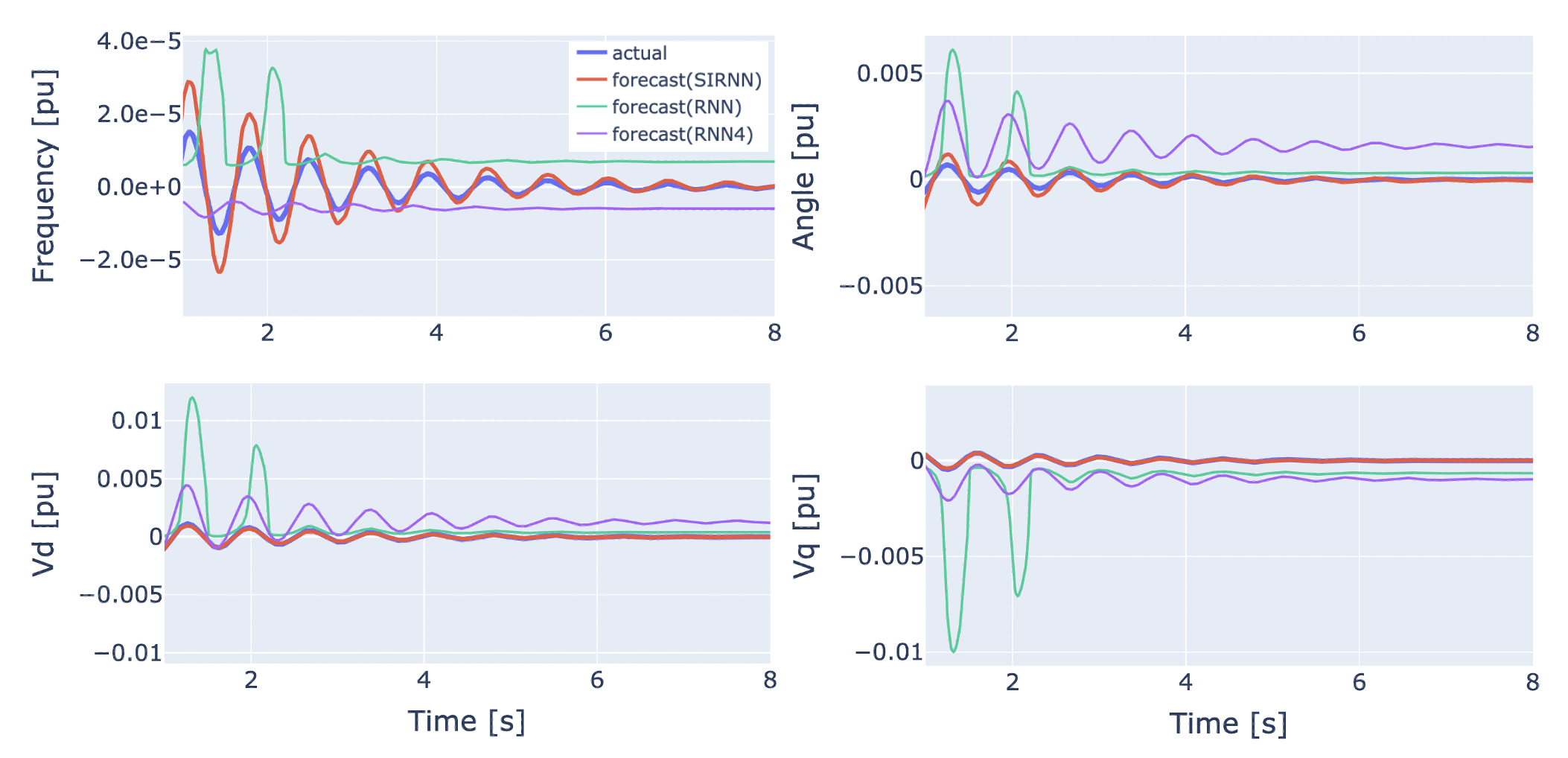}
\caption{ Comparisons of the proposed SI-RNN with RNN models on the SMIB test case in predicting the SG states for  one fault scenario from the test set. }
\label{fig:rnn_learning}
\end{figure*} 

We present the numerical results for the proposed SI-RNN-based dynamic learning approach, by considering both SGs and GFMs. To generate the datasets of dynamic trajectories, we use the time-domain (TD) simulations based on the $\text{Simulink}^{\text{\textregistered}}$ in $\text{Matlab}^{\text{\textregistered}}$. Both SGs and GFMs have been  represented using the three-phase full-order models, which are even more  complex than those in Sec.~\ref{sec:ps}. Particularly, the GFM model follows from~\cite{tayyebi2020ibr}, and has used the Electromagnetic Transient (EMT) simulation mode. To obtain the measurement data, we focused on educed-order states as in Remark~\ref{rmk:reduction}, and a smoothing and down-sampling process has been used to filter the internal states of dynamic components.  Additionally, the terminal bus voltage phasor data have been synthesized using the virtual Phasor Measurement Units (PMUs).

To evaluate the proposed SI-RNN's performance in predicting dynamic responses to disturbances, we simulate different fault scenarios as detailed in specific tests later. The dynamic sensitivity is numerically computed by using a step change in $P^\star$ after the fault clearance. The simulated trajectories for all the scenarios are split into the training/test datasets with a 90\%/10\% division. To better compare the performance, we consider two error metrics to evaluate the trained models on the scenarios in the test dataset. The validation error measures the deviations of NN outputs of one-step predictions from ground truth, while the prediction error is with respect to the trajectories of iterative predictions using only initial conditions. Hence, the latter should be slightly higher as it is prone to error propagation. As for the NN model setups, the proposed SI-RNN uses a total of four RNN blocks where the number of hidden features is 50 for each block, along with a final linear output layer. We also implemented the vanilla RNNs with the same parameter setup for comparisons. Both learning approaches have been implemented using the PyTorch library in Python.  They are trained by the standard AdamW algorithm with the same convergence criteria. The NVIDIA $\text{Quadro}^{\text{\textregistered}}$ RTX 5000 is utilized for computation acceleration\footnote{The simulation files, code, and results are available at: \newline \url{https://github.com/ShaohuiLiu/IBR_Dyn_Learn}}.


\subsection{Learning the SG model} \label{sec:sg}
We first validate the performance of the proposed SI-RNN for learning the full-order SG and controller models. A sixth-order SG model with a hydraulic turbine and governor, as well as an IEEE Type1 excitation system, is connected to an infinite bus through some transmission components to form a single machine infinite bus (SMIB) system, as depicted in Fig.~\ref{fig:SMIB_diagram}. A total of 400 fault scenarios are generated by changing the location of a three-phase grounded fault on the transmission line,  and duration from 1 to 10 cycles. For each scenario, the post-fault responses are simulated for 14 seconds with a 100Hz sampling rate. We compare the test performance of the proposed SI-RNN model with the vanilla RNN one in predicting the state trajectory. We use the performance metric of the normalized mean squared error (NMSE) given by
\begin{align}
 \mathrm{NMSE}=\frac{\| \bbx - \hhatbbx \|_2}{\| \bbx\|_2} 
\label{eq:est_err}
\end{align}
for predicting the ground-truth trajectory vector $\bbx=\{\bbx_t\}$. Examples of the test trajectories are presented in Fig.~\ref{fig:rnn_learning}, and the detailed comparison results are listed in Table~\ref{table:smib-accuracy}. 

\begin{figure}[t!]
\centering
\includegraphics[width=1\linewidth]{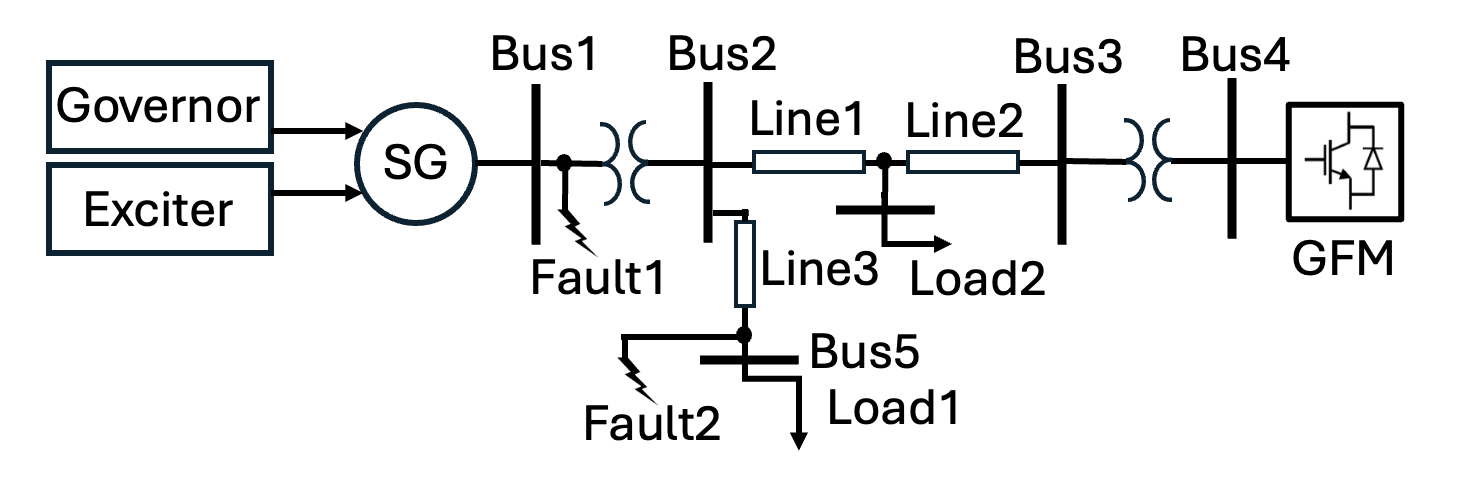}
\caption{ One-line diagram of the 5-bus test system with one SG and one GFM.}
\label{fig:GFM_diagram}
\end{figure} 

\begin{table}[t]
\centering
\caption{NMSE of predicting the SG states for the SMIB test system.} 
\begin{tabular}{c|ccc|ccc}

\toprule
\midrule
\multicolumn{1}{c}{ } &  \multicolumn{3}{c}{Validation error} & \multicolumn{3}{c}{Prediction error}\\
\midrule
State & RNN & RNN4 & \textbf{SI} & RNN & RNN4 & \textbf{SI} \\ 
\midrule
$\omega$ & 1.00    & 0.99 & \textbf{0.03}   & 1.04  & 1.00    & \textbf{0.07}   \\
$\delta$ & 0.24    & 0.81 & \textbf{0.04}    & 0.52  & 1.54    & \textbf{0.11}   \\
$E_\mathrm{d}$      & 0.32 & 0.67  & \textbf{0.04} & 0.50  & 1.28    & \textbf{0.11} \\
$E_\mathrm{q}$       & 0.38    & 0.80 & \textbf{0.06}    & 0.66  & 1.00    & \textbf{0.12}  \\  
\midrule
\bottomrule 
\end{tabular}
\label{table:smib-accuracy}
\end{table}

Fig.~\ref{fig:rnn_learning} and Table~\ref{table:smib-accuracy} show that both learning approaches can achieve good accuracy in predicting all the states in the validation test, while the SI-RNN slightly outperforms the RNN model. Nonetheless, when considering the iterative prediction using only the initialization, the proposed SI-RNN model exhibits much higher accuracy and numerical stability over the RNN model. Although adding additional historical trajectories will improve the accuracy of predicting oscillation modes as shown in Fig.~\ref{fig:rnn_learning}, the deviations caused by the instability will lead to larger prediction error as shown in the column `RNN4' in Table~\ref{table:smib-accuracy}. 
This comparison speaks of the SI-RNN's improved capability of tracking longer-term dependency and maintaining the stability. 

\subsection{Learning the GFM model}  

We further test the proposed SI-RNN for predicting the fast dynamics of IBRs, by using a 5-bus sub-system taken from the WECC 9-bus system. As shown in Fig.~\ref{fig:GFM_diagram}, there are 1 GFM and 1 SG, respectively connected to Bus4  and Bus1. The SG here has the same setup as in the previous SMIB system in Sec.~\ref{sec:sg}, while a GFM with dVOC as primary control has replaced the SG connected to Bus4 in the original 9-bus system.
A total of 200 fault scenarios have been simulated by using 2 different fault locations for a three-phase grounded fault of various fault durations from 1 to 10 cycles. For each scenario, the post-fault trajectories are generated by the EMT simulations for a time window of 3 seconds at a rate of 10kHz.  Note that the simulation time here is significantly shorter than that of the previous SMIB system with the single SG, and this is because the inclusion of the GFM has led to much faster damping of the system oscillations.
From the 10kHz EMT simulations, 
we further perform a filtering and down-sampling process to generate the measurement data at a rate of 100Hz for the states after reduction, as detailed in Remark~\ref{rmk:reduction}. 

\begin{figure*}[t!]
\centering
\includegraphics[width=1\linewidth]{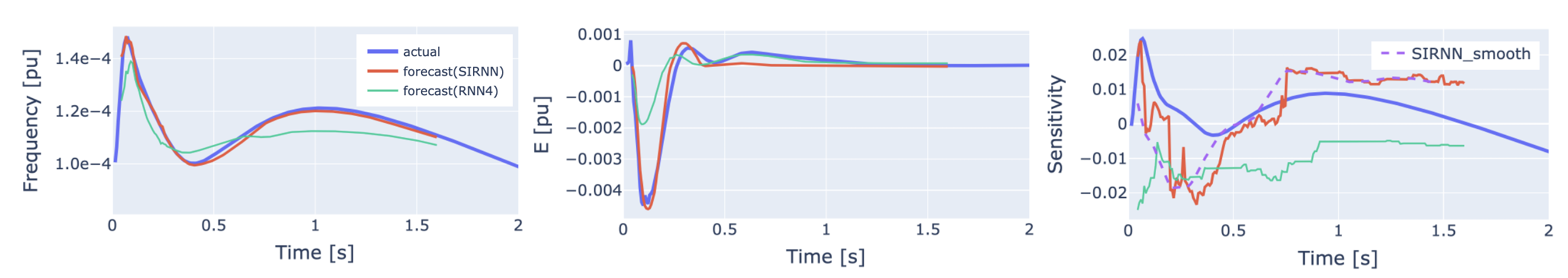}
\caption{ Comparisons of the proposed SI-RNN with RNN models on the 5-bus test case in predicting the GFM states  and $\omega-P_c$ dynamic sensitivity for one fault scenario from the test set. }
\label{fig:reduced_GFM_learning}
\end{figure*}


The validation test results on GFM state predictions are similar to the SG test case in the SMIB system, with SI-RNN outperforms the vanilla RNN, as shown in Fig.~\ref{fig:reduced_GFM_learning} and Table~\ref{table:2bus-accuracy}. Interestingly, the SI-RNN model have shown much higher accuracy and stability in iterative predictions over the vanilla RNN model for both SG and GFM states in the reduced-order model. This shows the capability of the proposed SI-RNN model in predicting nonlinear and fast dynamics. 


Using the GFM-based 5-bus system, we further compare the dynamic sensitivity predicted by the SI-RNN model with the numerical estimation method based on~\eqref{eq:data_sensitivity}. 
Fig.~\ref{fig:reduced_GFM_learning} plots the predicted versus the estimated sensitivity 
trajectories for a change of active power set-point $P^\star$ at the GFM right after the fault clearance.  Interestingly, the sensitivity in all scenarios from the test set are very similar to the numerical sensitivity calculated using simulated data. Nonetheless, the SI-RNN-based sensitivity trajectories are not smooth, which is consistent with the fact that the sensitivity is not time-dependent like states.

\begin{table}[t]
\centering
\caption{NMSE of predicting the states for the SG, and the GFM along with its trajectory sensitivity.} 
\begin{tabular}{c|cc|cc}

\toprule
\midrule
\multicolumn{1}{c}{ } &  \multicolumn{2}{c}{Validation error} & \multicolumn{2}{c}{Prediction error}\\
\midrule
SG & RNN & \textbf{SI-RNN} & RNN & \textbf{SI-RNN} \\ 
\midrule
$\omega$ & 0.78    & \textbf{0.05} & 0.81    & \textbf{0.17}   \\
$E$      & 0.76 & \textbf{0.05}  & 0.83 & \textbf{0.11} \\
\midrule
GFM & RNN & \textbf{SI-RNN} & RNN & \textbf{SI-RNN} \\ \midrule
$\omega$ & 0.04    & \textbf{0.04} & 0.13    & \textbf{0.11}   \\
$E$      & 0.12 & \textbf{0.05}  & 0.63  & \textbf{0.31} \\
Sensitivity  & 1.42 & \textbf{1.31}  & 2.03 & \textbf{1.16} \\
\midrule
\bottomrule 
\end{tabular}
\label{table:2bus-accuracy}
\end{table}



\section{Conclusions} 
\label{sec:con}

This paper develops a modularized learning framework to predict power system dynamics. We have effectively demonstrated a  reduced-order modeling approach to learn the dynamics of both SGs and IBRs as individual dynamic components, through a novel RNN-based framework that integrates high-order integral schemes to enhance stability and predictability. Our proposed SI-RNN model not only copes with the rapid dynamics characteristic of modern power systems but also ensures robust state sensitivity tracking in response to control signal changes. Numerical validations on datasets generated by full-order EMT simulations from the SMIB system and a 5-bus system demonstrate the validity and advantages of our proposed model over the regular RNN model, especially for predicting the GFM dynamics and the dynamic sensitivity. Numerical results demonstrate our approach paves the way for a flexible and efficient integration of all network-wide components in a modular fashion. 
Our ongoing works include coupling different types of modules and learn system-wide dynamics. Furthermore, we are actively investigating using predicted dynamic sensitivity to design learning-based control systems~\cite{gao2022inverse,issa2022artificial}.

\section*{Acknowledgement}
This work is partially supported by the NSF under Grants 2130706 and 2150571.
This work is partially supported by the U.S. Department of Energy’s Office of Energy Efficiency and Renewable Energy (EERE) under Solar Energy Technologies Office (SETO) Award Number 38637.

\balance

\section*{References}
\bibliography{IEEEabbrv,ref}
\bibliographystyle{IEEEtran}


\end{document}